\documentclass{article}

\usepackage[final,nonatbib]{nips_2018}

\usepackage[utf8]{inputenc} 
\usepackage[T1]{fontenc}    
\usepackage{hyperref}       
\usepackage{url}            
\usepackage{booktabs}       
\usepackage{amsfonts}       
\usepackage{nicefrac}       
\usepackage{microtype}      

\usepackage{graphicx}
\usepackage{subcaption}
\usepackage{amssymb}
\usepackage{url}

\title{AI Neurotechnology for Aging Societies \\ -- Task-load and Dementia EEG Digital Biomarker Development Using Information Geometry Machine Learning  Methods -- }

\author{
 Tomasz M. Rutkowski,  Qibin Zhao, Masato S. Abe, and Mihoko Otake\\
 RIKEN Center for Advanced Intelligence Project (AIP)\\
 Nihonbashi 1-chome Mitsui Building, 15th floor\\
1-4-1 Nihonbashi, Chuo-ku, Tokyo, 103-0027 Japan\\
\url{http://aip.riken.jp/} \\
 \texttt{tomasz.rutkowski@riken.jp}
 }

\begin{document}

\maketitle

\begin{abstract}\label{sec:intro}
Dementia and especially Alzheimer’s disease (AD) are the most common causes of cognitive decline in  elderly people. A spread of the above mentioned mental health problems in  aging societies is causing a significant medical and economic burden in many countries around the world. According to a recent World Health Organization (WHO) report, it is approximated that currently, worldwide, about 47 million people live with a dementia spectrum of neurocognitive disorders. This number is expected to triple by 2050, which calls for possible application of AI--based technologies to support an early screening for preventive interventions and a subsequent mental wellbeing monitoring as well as maintenance with so--called {\it digital--pharma} or {\it beyond a pill} therapeutical approaches. This paper discusses our attempt and preliminary results of brainwave (EEG) techniques to develop digital biomarkers for dementia progress detection and monitoring. We present an information geometry--based classification approach for automatic EEG--derived event related responses (ERPs) discrimination of low versus high task--load auditory or tactile stimuli recognition, of which amplitude and latency variabilities are similar to those in dementia. The discussed approach is a step forward to develop AI, and especially machine learning (ML) approaches, for the subsequent application to mild--cognitive impairment (MCI) and AD diagnostics.
\end{abstract}

\section{Introduction}

It is already a known fact that in the 21st century, dementia is the greatest global challenge for health and social care. Worldwide, mainly for people above 65 years old, dementia numbers and costs are rising due to increased longevity~\cite{lancet2017dementia}. Cabinet Office in Japan publishes annual reports on aging society to address the problem~\cite{agingJPgovREPORT}. United Nations Sustainable Development Goal~$\#3$ -- ``Good Health and Well--being"  focuses on healthy lives and it promotes well-being for all at all ages.
We propose an approach utilizing AI--based machine learning  for automatic discrimination of anomalous EEG brainwaves (ERPs), which shall lead to the development of digital biomarkers for task--load and dementia progress discrimination. The state--of--the--art methods for dementia diagnostics rely on standard subjective psychometric tests~\cite{moca2012}, or more contemporary behavioral evaluation within cognitive behavioral therapy (CBT) approaches such as a co--imagination method~\cite{otake2009coimagination}. Our proposal utilizes human noninvasive brainwave monitoring using EEG and event related potentials (ERPs) derived from responses to natural stimulus using modern comfortable recording setups~\cite{nozomuANDtomekAPSIPA2012,tomekFRONTIERS2016}. Methods developed by our team and presented in this paper shall lead to machine learning (ML) or AI in general applications allowing home--based monitoring of task--load or cognitive decline (dementia) with digital biomarkers. 
State--of--the--art research findings in neuroscience have illustrated that dementia brain decline is related to abnormal tau and amyloid disposal that modify pre-- and post--synaptic neuronal mechanisms, resulting in elevated neuronal calcium influxes causing an increased excitability, neuronal loss and finally altered brain rhythmic patterns~\cite{oscillopathy2006,theENDad2017,dementiaBIOMARKER2018review}. It is known that synaptic plasticity is essential for many cognitive functions (e.g. learning, abstract thinking, memory formation, etc.)~\cite{book:eeg}. 
In contemporary neurophysiological and pathological studies, dementia and especially AD have recently been identified as an impaired synaptic plasticity syndrome~\cite{dementiaBIOMARKER2018review}. 
The above shown age--related brain declines have also been characterized as network disconnection diseases and referred to as ``oscillopathies''~\cite{oscillopathy2006}. It has also been shown  that the so--called P300 EEG responses have significantly longer latencies, which, together with other ERP abnormalities, appear approximately $10$~years before dementia manifestation, especially in relation to deterioration of language, memory, and executive functions~\cite{LEE201362}. The dementia pre-- and post-- treatment P300 latency variabilities have also been correlated with cognitive ability and memory performance~\cite{dementiaBIOMARKER2018review}, thus we decided to focus on this particular response for the EEG biomarker development.
\begin{figure}[t]
	\centering
	\begin{subfigure}[b]{0.495\textwidth}
		\includegraphics[width=\textwidth]{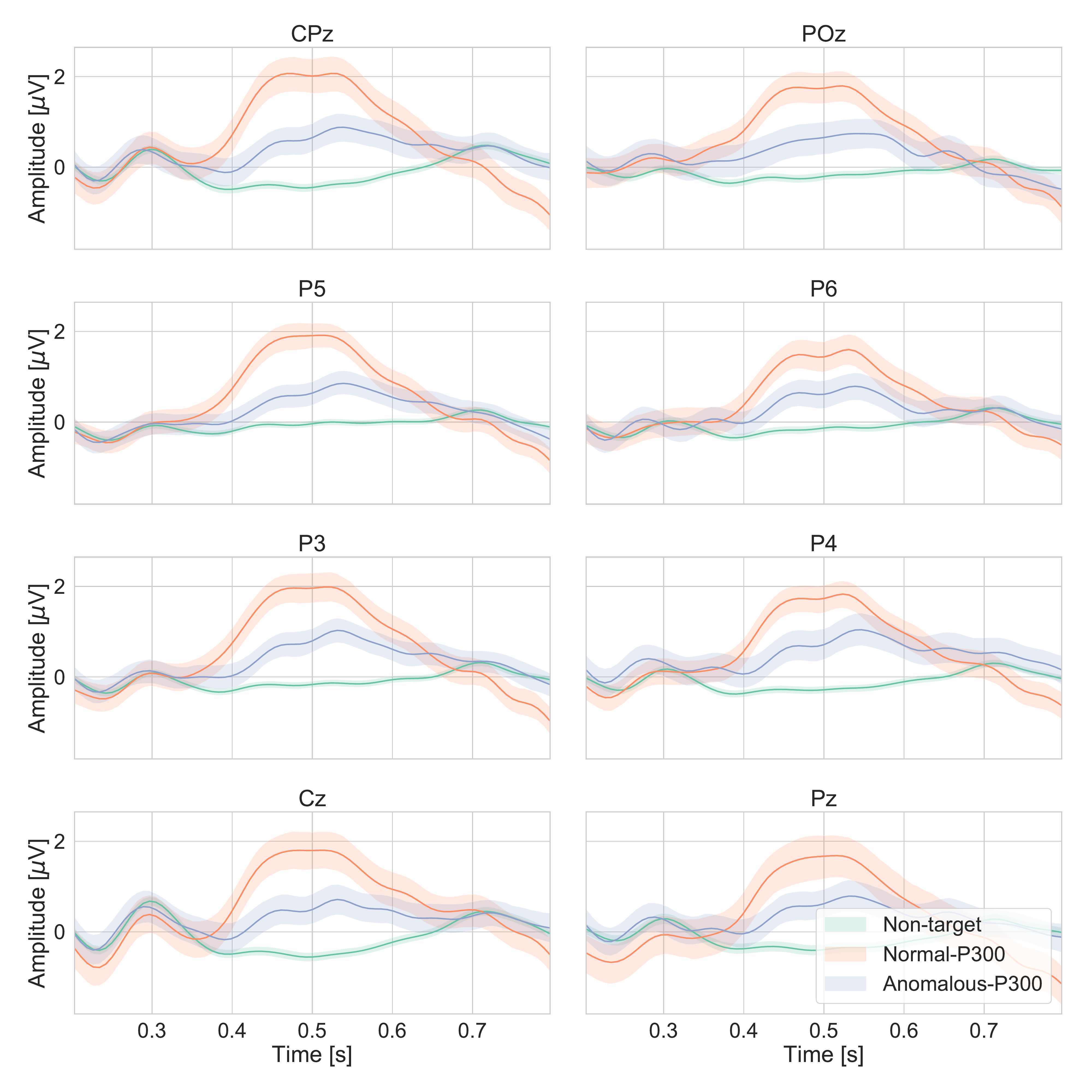} 
        		\caption{Amplitude modulated auditory P300 responses}
        		\label{fig:P300arv}
    	\end{subfigure}
    	\begin{subfigure}[b]{0.495\textwidth}
		\includegraphics[width=\textwidth]{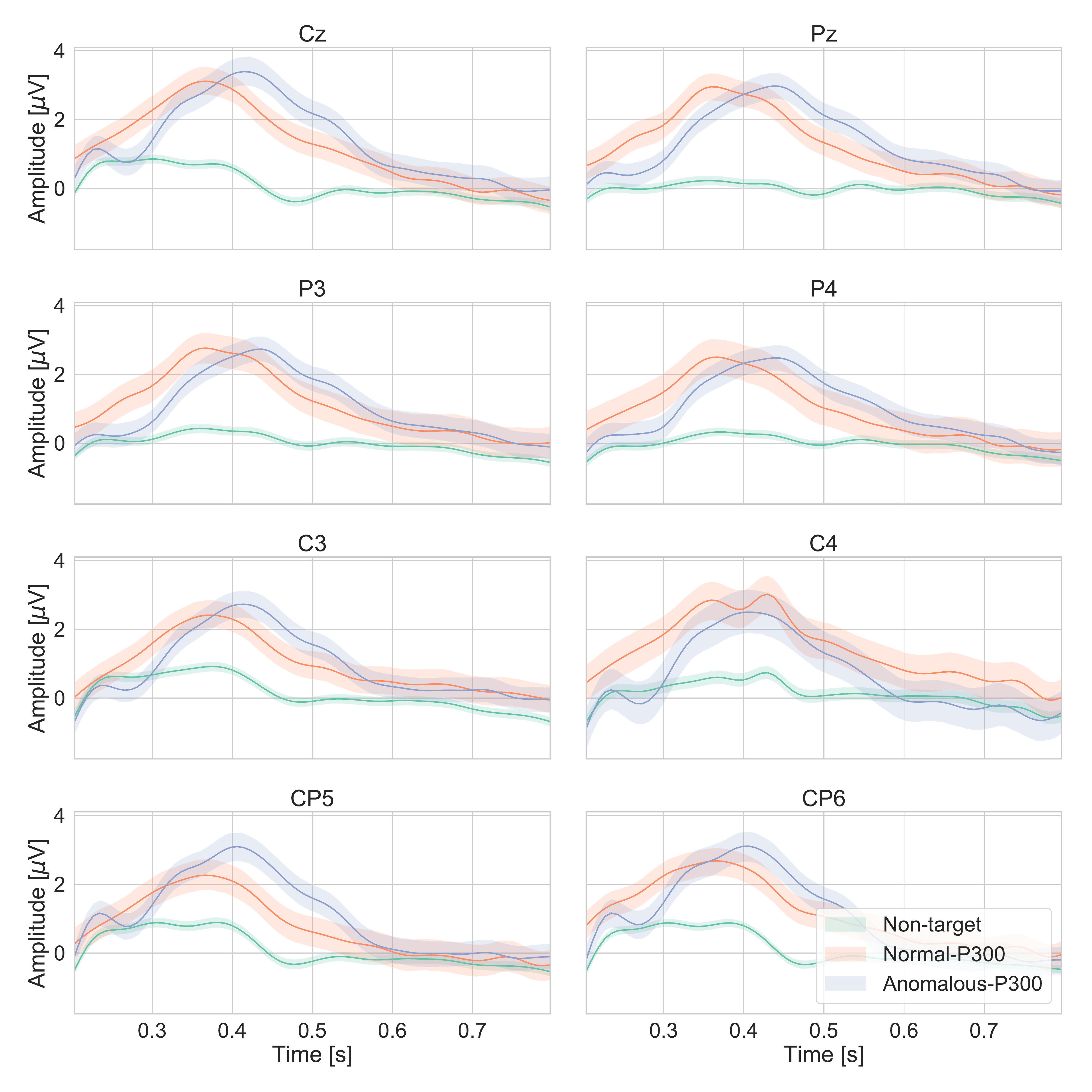} 
        		\caption{Time--shifted tactile P300 responses}
        		\label{fig:P300autd}
    	\end{subfigure}
	\caption{Averaged event--related potential responses (ERPs) as obtained from EEG measurements with nine and thirteen healthy subjects in auditory (\ref{fig:P300arv}) and tactile (\ref{fig:P300autd}) BCI experiments respectively. Eight EEG electrode traces are depicted with $95-$percentile error bars. Red traces correspond to easy (low task--load with normal--P300) responses to identified auditory or tactile targets with P300 responses, while blue ones correspond to the difficult (auditory virtual or tactile airborne ultrasonic with high task--load and anomalous--P300) responses, which model the dementia--based brainwaves~\cite{dementiaBIOMARKER2018review}. Green traces are the ignored (non--target) brainwaves without P300. In this paper we discriminate between the two P300--responses of low-- and high task--load for auditory (\ref{fig:P300arv}) and tactile (\ref{fig:P300autd}) cases to develop methods for subsequent application to the very similar dementia brainwaves~\cite{dementiaBIOMARKER2018review}.}\label{fig:p300}    
\vspace{-0.52cm}
\end{figure}
Recent approaches to dementia and AD resulted in a necessity to develop personalized therapies relying not only on traditional pharmacological interventions but also on lifestyle modifications~\cite{theENDad2017} as well as cognitive training~\cite{otake2009coimagination}. 
The classical pharmacological and the so--called {\it digital--pharma} or {\it beyond a pill} therapeutical approaches require trustful biomarkers, which would not only discriminate cognitive decline from resting state brainwave digital biomarkers~\cite{tomekICASSP2006} but all importantly would allow for continuous   home--based monitoring. We propose to utilize the so called task-- or work--load evaluation for dementia brainwave responses' discrimination in natural stimulation (auditory, visual, somatosensory, etc.) settings. 
In this paper we report on a development of EEG brainwave classification methods using dementia--modeling responses~\cite{oscillopathy2006,dementiaBIOMARKER2018review}. 
The spatial auditory~\cite{nozomuANDtomekAPSIPA2012} and airborne ultrasound tactile device (AUTD)~\cite{autdBCIconf2014} BCI paradigms  allow for a task--load modulation by incorporating real--sound/vibro--tactile or virtual--sound/AUTD stimuli, which are categorized as either easy or hard to perceive. 
The users are instructed to identify spatial targets (resulting with the so--called P300 responses) and ignore distractors as in the classical oddball paradigm experimental setup~\cite{nozomuANDtomekAPSIPA2012,tomekFRONTIERS2016,autdBCIconf2014,tomekJNM2015}. 
Averaged results with normal, anomalous (high task--load modeling dementia responses~\cite{oscillopathy2006,dementiaBIOMARKER2018review}) and ignored non--target ERPs are shown in Figure~\ref{fig:p300}.
We propose a framework to utilize AI--based neurotechnology employing ML together with recently proposed data--driven preprocessing techniques based on a  wavelet synchro--squeezing--transform (WSST)~\cite{tomekJNM2015,tomekJOMS2016} and Riemmanian geometry (RG) classification methods~\cite{barachant2010riemannian}.

\section{Methods}\label{sec:methods}

The brain responses used for EEG--based digital biomarker development in this study were collected from nine subjects in auditory, and thirteen in tactile, BCI projects approved by The Ethical Committee of Faculty of Engineering, Information and Systems at University of Tsukuba, Tsukuba, Japan.  
The currently presented EEG brainwave post--processing study was approved by The Ethical Committee of RIKEN, Wako--shi, Japan. 
At the EEG preprocessing stages we utilized a wavelet synchro--squeezing--transform (WSST) approach previously developed by the authors for BCI and sleep stages classification~\cite{tomekJNM2015,tomekJOMS2016}.
At the next step of the EEG processing pipeline we assumed that $x(t)\in\mathbb{R}$ was a zero--mean signal data sample captured from an EEG electrode at discrete time $t.$ 
In that case, let $\mathbf{X}_{k,i}\in\mathbb{R}^{N \times M}$ be an event $i$ representing an ERP response to a stimulus $k\in \{1,2\}$  from $N$ electrodes with $M$ samples. 
With an assumption of a zero mean, a sample covariance matrix of a given trial $\mathbf{X}_{k,i}$ belonging to a class $k$ has been given by $\mathbf{C}_{k,i} = \frac{1}{M-1}\mathbf{X}_{k,i} \mathbf{X}_{k,i}^T,$ as first proposed by~\cite{barachant2010riemannian}.
With an assumption that noisy EEG recordings have multivariate Gaussian distributions, a covariance matrix representing ERP features shall be considered as the only unique parameter for task--load or dementia stages. Features representing stimulus locked and segmented ERPs $k\in \{1,2\}$ were calculated  as $\mathbf{x}_{k,i}$ and transferred into covariance matrices $\mathbf{C}_{k,i}$. 
In a classifier training phase~\cite{barachant2014plug}, a geometric mean covariance matrix $\mathbf{\overline{C}}_{k}$ representing each ERP class $k$ was computed. In order to measure the distance of a newly recorded ERP to the above  class--representing mean matrix, the RG framework was used as proposed by~\cite{barachant2014plug}.  
A geodesic between two points $\mathbf{C}_i$ and $\mathbf{C}_j$ is the shortest path curve that connects them. The Riemannian distance between two covariance matrices could be computed as 
\begin{equation}
	\delta_R = \left|\left|\ln(\mathbf{C}_i^{-1}\mathbf{C}_j)\right|\right|_F = \sqrt{\sum_n[\ln(w_n)]^2},
\end{equation}
where $||\cdot||_F$ denotes a Frobenius norm and $w_1,\ldots,w_n$ the eigenvalues of $\mathbf{C}_i^{-1}\mathbf{C}_j$, respectively~\cite{barachant2014plug}. 
The geometric mean of $L$ covariance matrices representing a single ERP class was calculated as $D(\mathbf{C}_1,\cdots,\mathbf{C}_l) = \arg\min_{\mathbf{C}}\sum_{l=1}^L\delta_R^2(\mathbf{C},\mathbf{C}_l).$ The geodesic on the manifold, according to the RG principles~\cite{barachant2014plug} was obtained from  
	$\Gamma(\mathbf{C}_i,\mathbf{C}_i, \tau) = \mathbf{C}_i^{\frac{1}{2}}\left(\mathbf{C}_i^{-\frac{1}{2}}\mathbf{C}_j\mathbf{C}_i^{-\frac{1}{2}}\right)^\tau\mathbf{C}_i^{\frac{1}{2}},$
with a scalar $\tau\in\{0,1\}$.
A frequently used classification approach for RG--based features has been based on a distance evaluation between the newly arriving ERP feature and mean covariance matrices~\cite{barachant2010riemannian,barachant2014plug} representing desired classes $k$. A suitable approach for the above procedure has been based on a minimum distance to mean (MDM) classifier~\cite{barachant2010riemannian,barachant2014plug}, which satisfied the above criterium. 
The MDM approach was very generic and easy to apply. We compared it with classical vectorized time domain ERP features~\cite{tomekJNM2015}; spatial filter xDAWM preprocessing~\cite{hiroshiAPSIPA2016}; tangent space mapping of RG features~\cite{barachant2014plug}; using linear regression (LR); regularized linear discriminant analysis (rLDA); linear, sigmoid and radial basis function support vector machine (linearSVM, sigmSVM and rbfSVM) methods as shown in Figure~\ref{fig:acc}.

\section{Results}
\label{sec:results}
 
\begin{figure}
	\centering
  	\begin{subfigure}[t]{0.48\textwidth}
		\includegraphics[width=\textwidth]{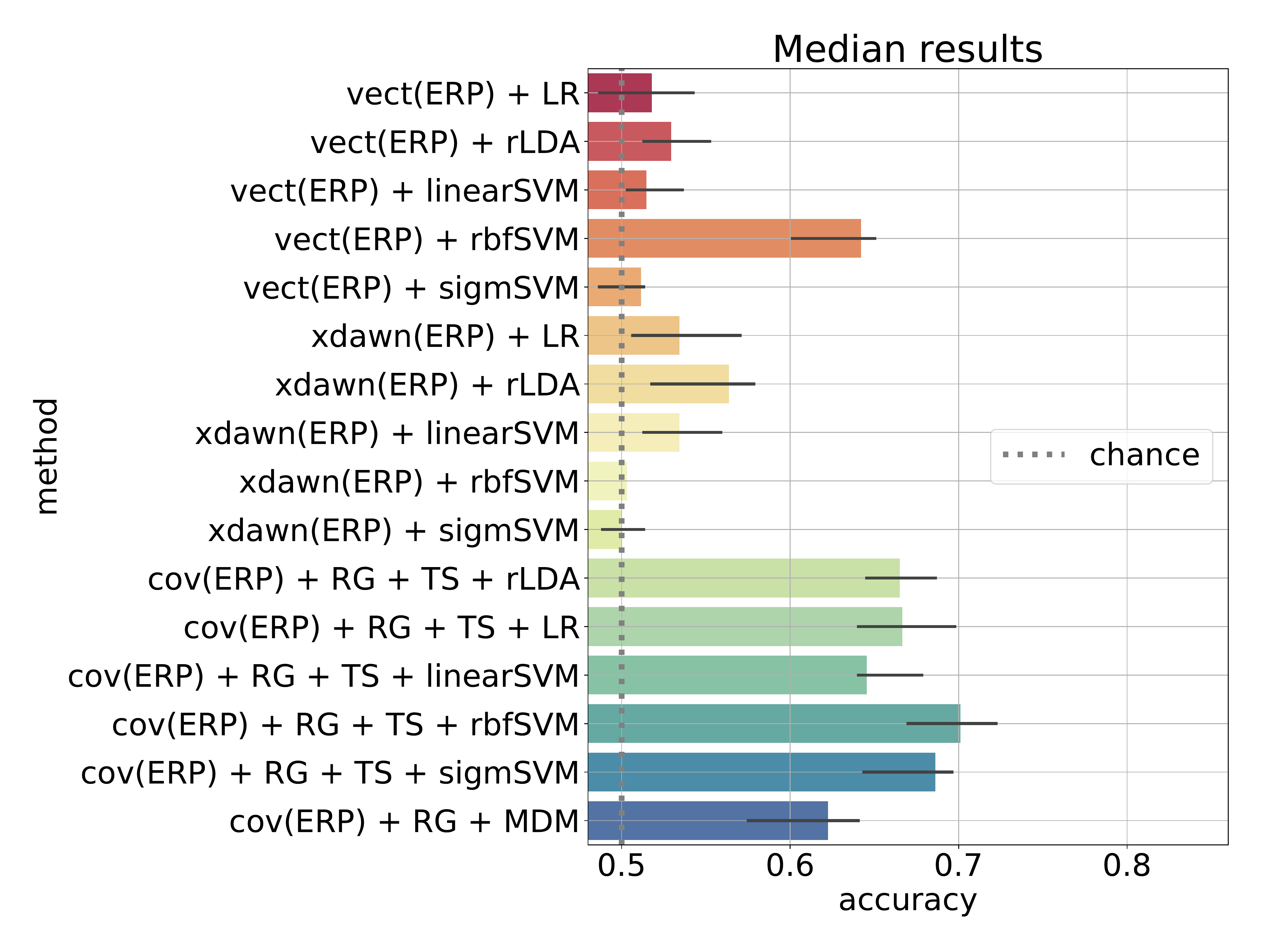} 
        		\caption{Median cross--validation accuracies for all subject dataset of amplitude modulated auditory (real versus virtual sound sources) P300 responses}
        		\label{fig:P300arv_all}
    	\end{subfigure}
	~
    	\begin{subfigure}[t]{0.48\textwidth}
		\includegraphics[width=\textwidth]{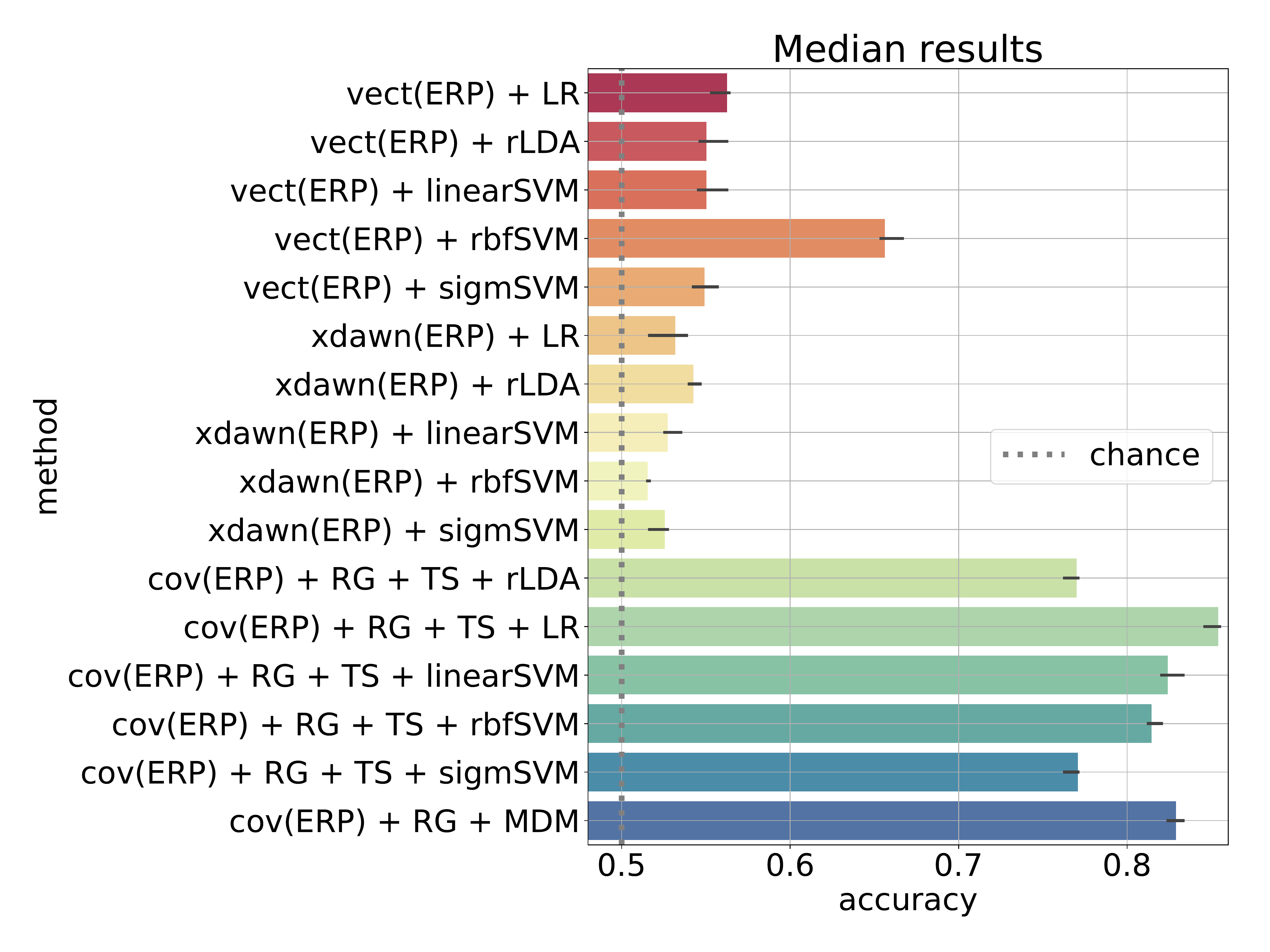} 
        		\caption{Median cross--validation accuracies for single subject datasets of amplitude modulated auditory auditory (real versus virtual sound sources) P300 responses}
        		\label{fig:P300arv_single}
    	\end{subfigure}
	\\ 
    	\begin{subfigure}[t]{0.48\textwidth}
		\includegraphics[width=\textwidth]{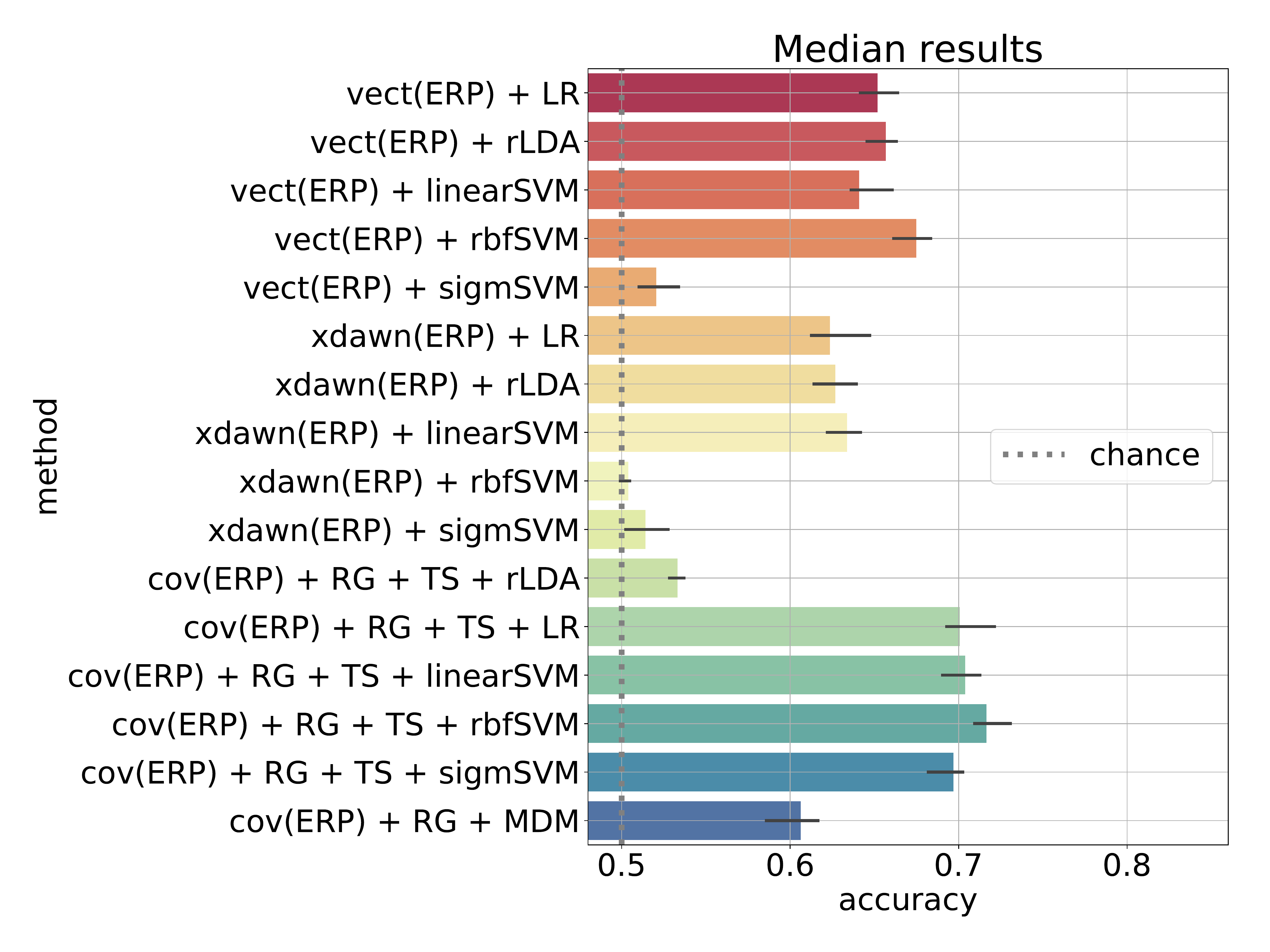} 
        		\caption{Median cross--validation accuracies for all subject dataset of time--shifted tactile (vibro--tactile versus AUTD stimuli) P300 responses}
        		\label{fig:P300autd_all}
    	\end{subfigure}
	~
    	\begin{subfigure}[t]{0.48\textwidth}
		\includegraphics[width=\textwidth]{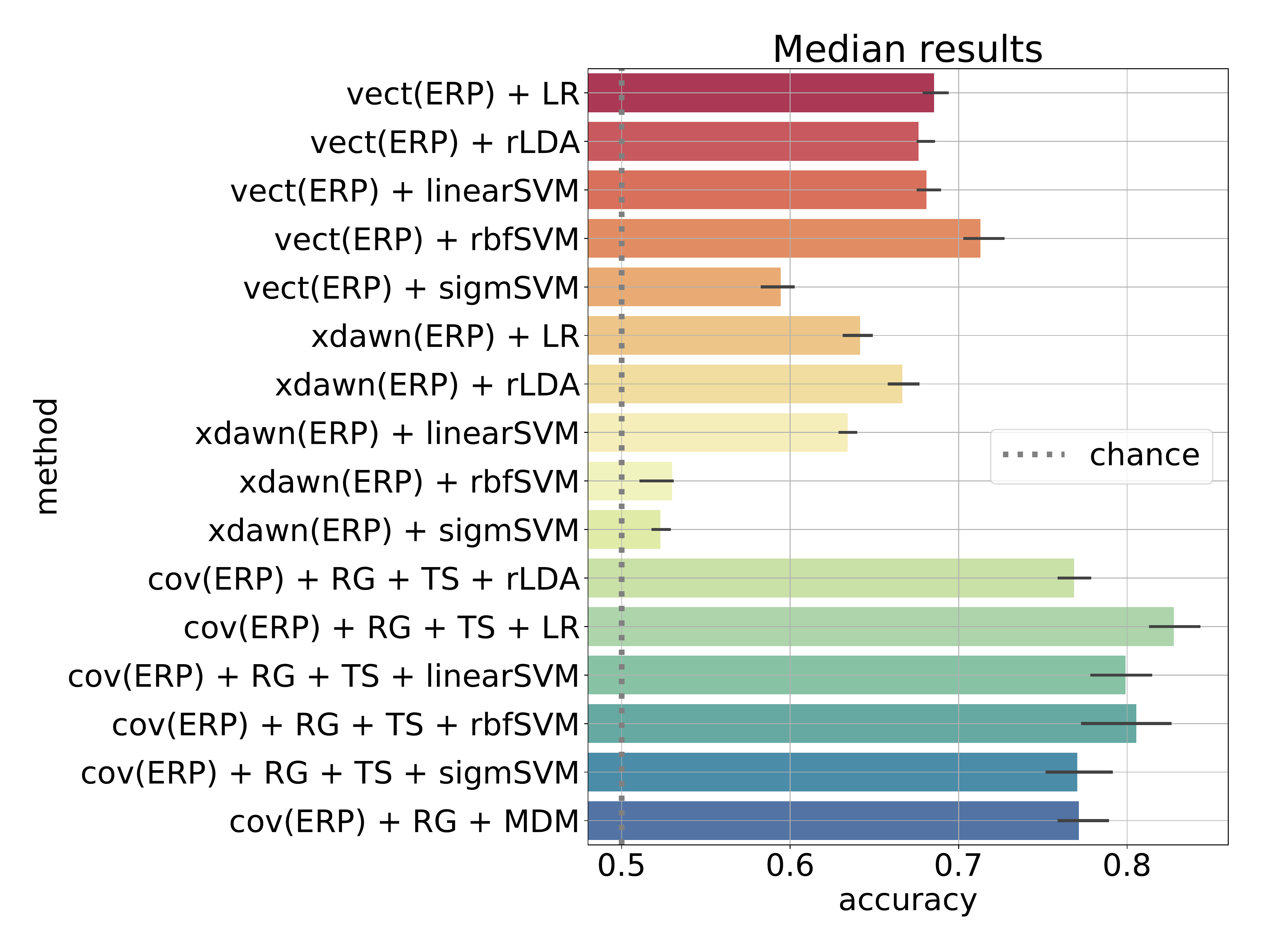} 
        		\caption{Median cross--validation accuracies for single subject datasets of time--shifted tactile P300 responses}
        		\label{fig:P300autd_single}
    	\end{subfigure}
	\caption{Median cross--validation accuracy results with a chance level of $0.5$  depicted with dotted gray line for amplitude modulated auditory  (\ref{fig:P300arv_all} and \ref{fig:P300arv_single}) as well as tactile (\ref{fig:P300autd_all} and \ref{fig:P300autd_single}) in all subject transfer learning (\ref{fig:P300arv_all} and \ref{fig:P300autd_all}) versus single user (\ref{fig:P300arv_single} and \ref{fig:P300autd_single}) datasets. }
	\label{fig:acc} 
	\vspace{-0.5cm}   
\end{figure}

The AI--neurotechnology--based approach using machine learning for task--load identification, which has been proposed as a model for dementia cognitive responses, resulted in a very encouraging classification boost for single user and transfer learning (a classifier training using brainwaves from many subjects).  
A very solid and statistically significant improvement of automatic task--load--dependent ERP classification was obtained. 
We obtained results from the nine and thirteen subject datasets  using the single user (see Figures~\ref{fig:P300arv_single}~and~\ref{fig:P300autd_single})
and transfer learning (see Figures~\ref{fig:P300arv_all}~and~\ref{fig:P300autd_all}) approaches with classifiers' cross--validated using class--balanced training sets.
The results are summarized in Figure~\ref{fig:acc} in the form of cross--validated ($10\%$ test data) accuracy distribution plots.
The best results (see Figure~\ref{fig:acc}) were obtained for tangent space mapping--based LR and SVM applications in single user and transfer learning training scenarios, comparing to the MDM as well as to the simple vectorized time domain ERP features (except radial basis function SVM for the latter cases). The majority of differences were statistically significant in pairwise comparisons (except the auditory case of radial basis function SVM versus TS RG linear SVM) as tested with non--parametric Wilcoxon pairwise tests at a level of $p < 0.05.$ Results of between-subject generalization trials (testing of classifiers using an external subject's EEG) were non--significantly different from the chance level. 

\vspace{-0.05cm} 
\section{Conclusions}
\label{sec:conclusions}
\vspace{-0.05cm} 

This study confirmed a successful application of the data--driven and information geometry--based automatic task--load classification approach as evaluated with nine and thirteen healthy subject datasets. The proposed approach resulted in a satisfactory classification (above $70\%$) of two task--load cases, namely easy versus difficult auditory and tactile paradigms, as shown in the form of averaged ERP traces in Figure~\ref{fig:p300}, and final results in Figure~\ref{fig:acc}.
The presented results of the automatic brainwave--based task--load classification offer a step forward in the development of novel dementia--related biomarkers for elderly people life improvement and healthcare cost lowering. The task--load ERP responses are very similar to those in dementia cases and they suggest a possibility to develop personal longitudinal biomarkers for cognitive wellbeing monitoring over time, as the obtained single user results have been the most satisfactory.
In the next step of our research project we plan to evaluate the developed methods with elderly normal versus dementia (subjective or mild cognitive impairment, etc.) diagnosed subjects, as well as using deep learning models. 

\small

\end{document}